\documentclass{cseg2014}
\usepackage{xy}
\usepackage{amsmath,graphicx,natbib}
\usepackage{rotating}
\usepackage{lscape}
\usepackage{color}

\begin{document}
\begin{center}
\vspace{30pt}
\includegraphics[]{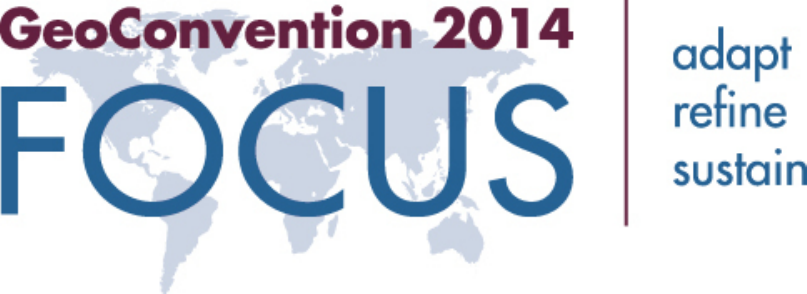}
\end{center}

\Large \color{black}{\bf Statics Preserving Sparse Radon Transform}\\
\small \color{black}\emph{Nasser Kazemi, Department of Physics, University of Alberta,
kazemino@ualberta.ca}

\section{Summary}
 This paper develops a Statics Preserving Sparse Radon transform (SPSR) algorithm.The de-coloration power of Radon basis functions depends on different factors. The most important one is statics. Statics decrease the sparsity of Radon models. To tackle this problem, it is necessary to include statics into the Radon bases functions. SPSR algorithm includes statics into the Radon bases functions, allowing for a sparse representation of statics contaminated data in the Radon domain. SPSR algorithm formulated as an $L_{2}- L_{1}$ optimization problem and solved via alternating minimization method. Real data examples used to test the performance of the proposed method in multiple and noise attenuation in the presence of statics. 
\section{Introduction}
By now it is well known that Radon transform operator is not orthogonal. Forward and inverse transformation without losing the data is not a trivial task. \cite{Hampson} and \cite{Beylkin} proposed a damped least squares solution to tackle the problem. However, this method suffers from the lack of resolution in the Radon domain.

Resolution of the Radon models depend on several factors. Limited aperture of the data, missing traces, aliasing, finite number of scanned ray parameter, noise and statics are the major sources of artifacts. This fact implies that Radon basis functions are not complete and components in the complimentary space of the Radon domain will be zeroed out. The harm effect of all  the aforementioned factors but statics can be compensated by constraining the solution space and searching for sparse models.To increase the resolution of Radon transform, \cite{Sacchi} introduced a sparse Radon algorithm in the frequency domain. They used Cauchy norm to model the probability density function. The major disadvantage of this method is that one need to assign deferent hyper-parameters for each of the frequency components. For more information interested readers can see \cite{Trad} and references therein.  On the other hand, statics is a serious problem in seismic processing. Statics reduce the efficiency of the methods that deal with spacial coherency and predictability of the data \cite[]{Traonmilin}. Hence, to improve the efficiency of the classical processing techniques, statics should be removed from the data. \cite{Stanton} and \cite{Gholami} showed that sparsity maximization can be used as a criterion for short period statics correction.   

In this paper I will show that short period statics drastically decreases the resolution of radon model. An algorithm will be introduced that simultaneously estimates the static shifts and yields high resolution Radon model. Moreover, the efficiency of the proposed method in noise and multiple suppression will be examined through real data example.

\section{Theory}
In this section I will explain SPSR method step by step. Data generating model under the action of the Radon forward operator $\bf{L}$ can be written as follow:
\begin{equation}\label{n1}
\bf{d}=\bf{L} \bf{m}+\bf{n},
\end{equation}
where $\bf{d}$ is data, $\bf{m}$ is radon model and $\bf{n}$ is the noise content.\cite{Thorson} showed that given the data, equation (\ref{n1}) can be solved via damped least squares approach. In other words, the objective function is
\begin{equation}\label{n:2}
\mbox{minimize}\;  ||{\bf m}||_2\;\;\;\;\;\;\;\;\;\;\;\;\;\;\;\;\;\;\;\;\;\;\;
\mbox{s.t.}\; ||{\bf d}-\bf{L} {\bf m}||_2< \epsilon
\end{equation}
where $\epsilon$ is some estimate of noise level in the data. To increase the resolution of the Radon model, one can adopt ${\emph{l}}_2$ norm for data misfit and ${\emph{l}}_1$ norm for the model 
\begin{equation}\label{n:3}
\widehat{{\bf m}}=\underset{{\bf m}}{\operatorname{argmin}}
\quad\quad\frac {1}{2}\|{\bf d}-{\bf L}{\bf m}\|_2+\tau\;\|{\bf{m}}\|_1,
\end{equation} 
where ${\bf m}$ is desired sparse model and $\tau$ is well known as regularization parameter that balances the importance of misfit functional and regularization term. The cost function of equation (\ref{n:3}) is complete if data has no statics. As discussed earlier statics introduce lots of artifacts and smear the Radon model. Considering this fact into account changes the equation (\ref{n:3}) to  
\begin{equation}\label{n:4}
\underset{{\mathbf m,Q}}{\operatorname{argmin}}
\quad\quad\frac {1}{2}\|{\bf Qd}-{\bf L}{\bf m}\|_2+\tau\;\|{\bf{m}}\|_1,
\end{equation} 
where $\bf{Q}$ is shifting operator that corrects for statics. Equation (\ref{n:4}) can be rewritten as follow
\begin{equation}\label{n:5}
\underset{{\mathbf m,Q}}{\operatorname{argmin}}
\quad\quad\frac {1}{2}\|{\bf d}-{\bf Q}^{T}{\bf L}{\bf m}\|_2+\tau\;\|{\bf{m}}\|_1,
\end{equation} 
where ${\bf{Q}}^{T}$ is the adjoint operator of $\bf{Q}$ and puts statics back into the Radon predicted data. The only advantage in using equation (\ref{n:5}) instead of equation (\ref{n:4}) is that equation (\ref{n:5}) preserves statics in the predicted data.
The cost function of equation (\ref{n:5}) can be minimized by alternatively solving the following sub-problems:
\begin{equation}\label{n:6}
\mbox{{\bf{m}}- step}\quad\quad \widehat{{\bf m}}= \underset{{\mathbf m}}{\operatorname{argmin}}
\quad\quad\frac {1}{2}\|{\bf{d}}-\widehat{{\bf Q}}^{T}{\bf L}{\bf m}\|_2+\tau\;\|{\bf m}\|_1,
\end{equation}
\begin{equation}\label{n:7}
\mbox{{\bf{Q}}- step}\quad\quad  \widehat{\bf{Q}}= \underset{{\mathbf Q}}{\operatorname{argmin}}
\quad\quad\frac {1}{2}\|{\bf d}-{\bf Q}^{T}{\bf L}\widehat{{\bf m}}\|_2.
\end{equation}

By considering the initial shifting operator as an identity matrix, the $\mbox{{\bf{m}}- step}$ can be solved using Fast Iterative Shrinkage-Thresholding Algorithm \cite[]{Beck}. The $\mbox{{\bf{Q}}- step}$ solved via maximum likelihood estimator of shifts which is  simply cross correlating the ${\bf{d}}$ and ${\bf{L}\widehat{\bf{m}}}$ vectors. Note, this is also a valid solution if we define the ${\bf{Q}}$ operator space as a combination of some shifting basis functions. 

In next section I will examine the efficiency of the proposed method in noise and multiple suppression using real data examples. 
\section{Examples}
\begin{figure}
\centerline{\includegraphics[width=.97\columnwidth]{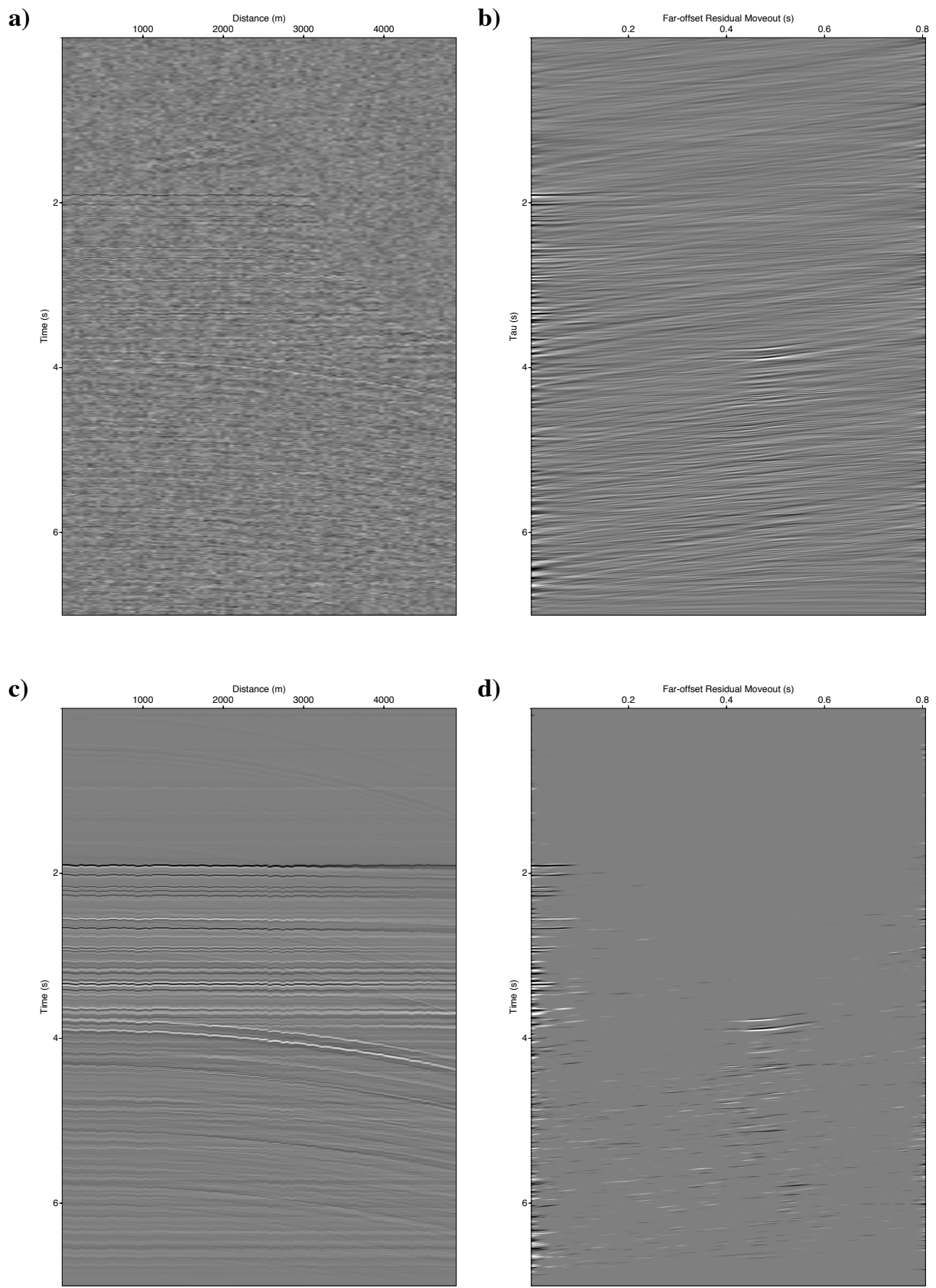}}
\caption{Comparison of conventional Radon transform and SPSR method. a) NMO corrected gather with noise and statics. b) Parabolic radon transform of data in a). c) SPSR estimated data with preserving statics. d) SPSR model of data in a).  }
\label{fig:1}
\end{figure}

To test the performance of the proposed algorithm I show an NMO corrected CMP gather from a Gulf of Mexico towed-streamer dataset. I applied random statics with maximum $\pm 10\;(ms)$ time shift. Moreover, I added a band limited Gaussian noise to the gather (Figure \ref{fig:1}a). As discussed earlier events in CDP gathers after NMO correction can be modelled as a parabola. Hence, I applied parabolic Radon transform. Figure \ref{fig:1}b shows the parabolic Radon model of the gather in Figure \ref{fig:1}a. It is obvious that because of the added noise, limited aperture and statics there are lots of artifacts and smearing in the model. Figures \ref{fig:1}c and d show the SPSR predicted data with preserved statics and SPSR model, respectively.The desired signal (residual move- out information) is easily recognizable in the SPSR model. It is also possible to apply filtering and remove multiples from the predicted data.

\begin{figure}
\centerline{\includegraphics[width=.97\columnwidth]{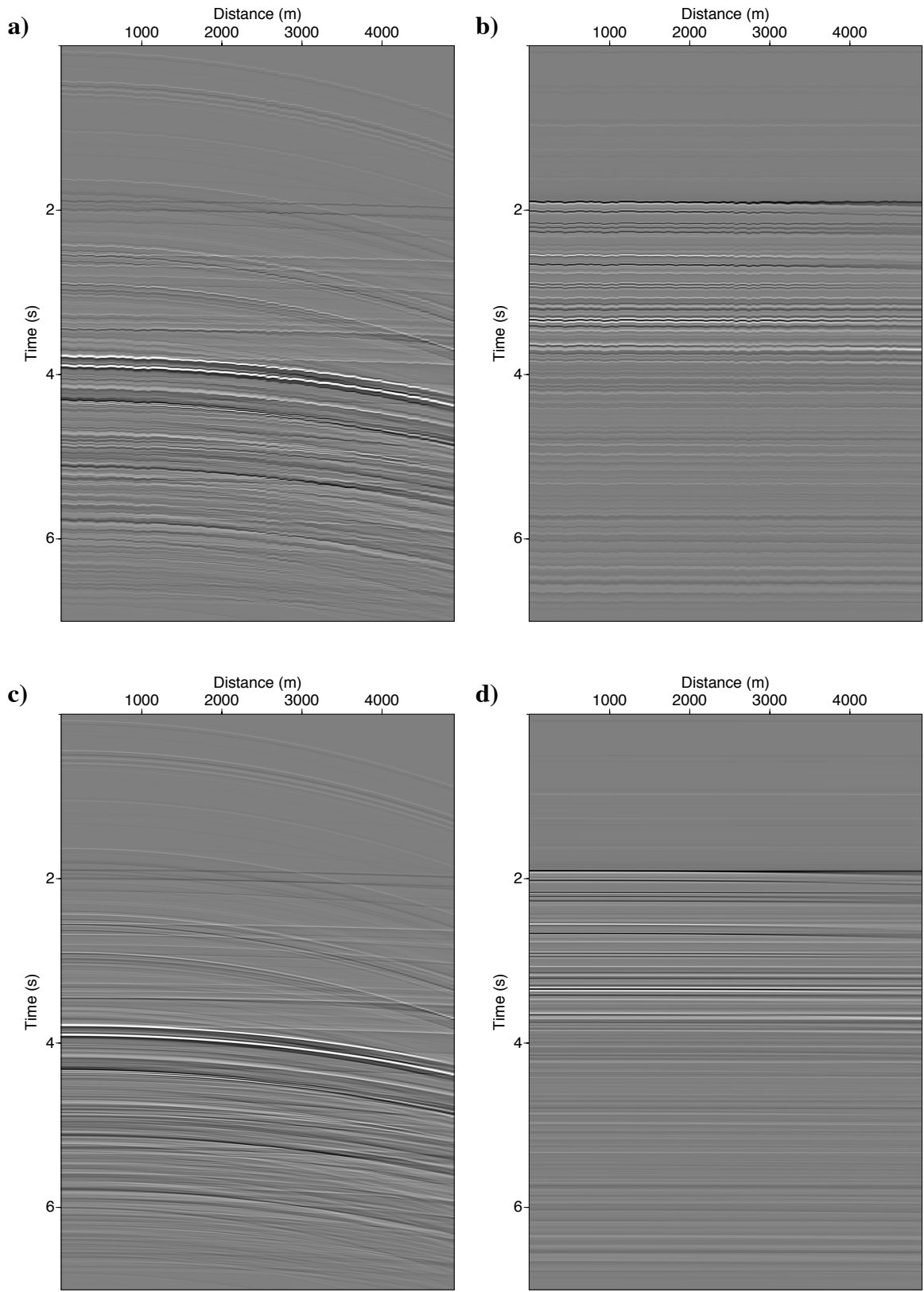}}
\caption{Primaries and multiples estimation using SPSR method with and without statics preserving. a) Estimated multiples with preserving statics. b) Estimated primaries with preserving statics. c) Estimated multiples after removing statics. d) Estimated primaries after removing statics.  }
\label{fig:3}
\end{figure}

It is well understood that multiples after NMO correction with primary events velocity show an under correction behaviour. In other words, primaries and multiples are separated in the Radon domain based on residual move-out. By muting zero and close to zero residual move-out region in the SPSR model and mapping the muted model to the data domain one can predict multiples. Figures \ref{fig:3} a and b show the estimated statics preserved multiples and primaries using the filtered version of SPSR model in Figure \ref{fig:1}d, respectively. Figures \ref{fig:3} c and d show the estimated multiples and primaries after removing statics. The SPSR algorithm did a good job in estimating the statics shifts and it is also yields hight resolution Radon model.

\section{Conclusions}
Radon transform is a powerful tool in seismic data processing. However, it suffers from variety of artifacts. The most important source of artifact is statics. It is shown that, statics decrease the de-coloration power of Radon basis functions. Hence, removing statics is a necessary step in improving the resolution of Radon transform. In this paper a statics preserving sparse radon transform method proposed for simultaneously predicting the statics in the data and increasing the resolution of Radon model. The tests suggest that the proposed method can be a practical tool for noise and multiple suppression of the statics contaminated datasets.
\section{Acknowledgments}
The author thank the sponsors of the Signal Analysis and Imaging Group (SAIG) at the University of Alberta.

\bibliographystyle{seg} 
\bibliography{CSEGpaper2014}

\begin{thebibliography}{}
\itemsep0pt

\bibitem[Beck and Teboulle, 2009]{Beck}
Beck, A. and M. Teboulle,  2009, A fast iterative shrinkage-thresholding
  algorithm for linear inverse problems: SIAM J. Img. Sci.,  183--202.

\bibitem[Beylkin, 1987]{Beylkin}
Beylkin, G.,  1987, Discrete radon transform: IEEE Trans. Acoust., Speech, and
  Sig. Proc, {\bf 35}, 162--172.

\bibitem[Gholami, 2013]{Gholami}
Gholami, A.,  2013, Residual statics estimation by sparsity maximization:
  Geophysics, {\bf 78}, V11--V19.

\bibitem[Hampson, 1986]{Hampson}
Hampson, D.,  1986, Inverse velocity stacking for multiple elimination: J. Can.
  Soc. Expl. Geophys, {\bf 22}, 44--55.

\bibitem[Sacchi and Ulrych, 1995]{Sacchi}
Sacchi, M. and T. Ulrych,  1995, High-resolution velocity gathers and offset
  space reconstruction: Geophysics, {\bf 60}, 1169--1177.

\bibitem[Stanton et~al., 2013]{Stanton}
Stanton, A., N. Kazemi, and M. Sacchi,  2013, Processing seismic data in the
  presence of residual statics: SEG Technical Program Expanded Abstracts,
  1838--1842, doi:10.1190/segam2013--1453.1.

\bibitem[Thorson and Claerbout, 1985]{Thorson}
Thorson, R. and J. Claerbout,  1985, Velocity-stack and slant-stack stochastic
  inversion: Geophysics, {\bf 50}, 2727--2741.

\bibitem[Trad et~al., 2003]{Trad}
Trad, D., T. Ulrych, and M. Sacchi,  2003, Latest views of the sparse radon
  transform: Geophysics, {\bf 68}, 386--399.

\bibitem[Traonmilin and Gulunay, 2011]{Traonmilin}
Traonmilin, Y. and N. Gulunay,  2011, Statics preserving projection filtering:
  SEG Technical Program Expanded Abstracts, {\bf 30}, 3638--3642.

\end{thebibliography}

\end{document}